\documentclass[aps,pra,twocolumn,10pt,letterpaper,superscriptaddress,showpacs,floatfix]{revtex4-2}

\usepackage[utf8]{inputenc}
\usepackage{CJKutf8}
\usepackage[T1]{fontenc}
\usepackage{microtype}
\usepackage{amsmath}
\usepackage{amssymb}
\usepackage{amsthm}
\usepackage{graphicx}
\usepackage{braket}
\usepackage{hyperref}
\usepackage{xcolor}
\usepackage{bm}
\usepackage{ragged2e}

\newtheorem*{theorem*}{Theorem}

\begin{document}

\begin{CJK*}{UTF8}{bsmi}

\title{A universal description of Mott insulators:\texorpdfstring{\\}{ } Characterizing quantum phases beyond broken symmetries}

\author{Matheus de Sousa}
\altaffiliation{These authors contribute equally}
\affiliation{School of Physics and Astronomy, Shanghai Jiao Tong University, Shanghai 200240, China}

\author{Zhiyu Fan}
\altaffiliation{These authors contribute equally}
\affiliation{School of Physics and Astronomy, Shanghai Jiao Tong University, Shanghai 200240, China}

\author{Wei Ku (\CJKfamily{bsmi}顧威)}
\altaffiliation{email: weiku@sjtu.edu.cn}
\affiliation{School of Physics and Astronomy, Shanghai Jiao Tong University, Shanghai 200240, China}

\date{\today}

\begin{abstract}
Using Mott insulators as a prototypical example, we demonstrate a \textit{dynamics}-based characterization of quantum phases of matter through a general $N$-body renormalization group framework.
The essential ``Mott-ness'' turns out to be characterized by a change of size-scaling of the \textit{effective} intra-\textit{momentum} repulsions between long-lived emergent ``eigen-particles'' that encodes the dynamics of two-body bound states in the high-energy sector.
This directly offers a \textit{universal} characterization at long space-time scale for the corresponding class of Mott insulators through a uniform single occupation of all momenta, and otherwise Mott metals.
This universal description naturally paves the way to topological Mott insulators and is straightforward to extend to bosonic Mott systems.
More generally, this demonstration exemplifies a generic paradigm of characterizing quantum phases of matter through their distinct \textit{dynamics} beyond broken symmetries.
\end{abstract}

\maketitle
\end{CJK*}

\textit{Introduction} —
Mott insulators represent a paradigmatic example of strongly correlated electron systems, with insulating behavior (unexpected from standard band filling considerations) emerges from strong electron-electron interactions~\cite{boerSemiconductorsPartiallyCompletely1937,mottDiscussionPaperBoer1937,mottBasisElectronTheory1949}.
Such interaction-induced insulators have been observed and intensively studied among a wide variety of material families, from transition metal oxides~\cite{cyrotTheoryMottTransition1972,terakuraTransitionMetalMonoxidesBand1984,anisimovBandstructureDescriptionMott1990}, rare-earth nickelates~\cite{Torrance1992, Medarde1997, Catalano2018}, layered Ruthenates like $\mathrm{Ca_2RuO_4}$~\cite{Nakatsuji_1997}, to organic charge-transfer salts such as $\mathrm{(BEDT-TTF)_2X}$~\cite{PhysRevLett.91.016401,PhysRevB.69.064511,Powell_2006}, all displaying interesting magnetic~\cite{anderson1950antiferromagnetism,kugel1982jahn}, orbital~\cite{tokura2000orbital,khaliullin2005orbital}, and lattice~\cite{Zhang_2019,Natori_2019} properties.
Particularly in vanadium oxide $\mathrm{V_2O_3}$, a paradigmatic metal-insulator transition can be realized through modulation of temperature, and pressure, making it a prototypical system for studying Mott physics~\cite{mottMetalinsulatorTransitionExtrinsic1972,mottMetalinsulatorTransitionsVO21974,zylbersztejnMetalinsulatorTransitionVanadium1975,hansmannMottHubbardTransitionV2O32013}.

Furthermore, upon introduction of additional carriers through chemical doping, most of the doped Mott insulator such as Cuprates $\mathrm{La_2CuO_4}$~\cite{bednorzPossibleHighTcSuperconductivity1986,leeDopingMottInsulator2006,keimerQuantumMatterHightemperature2015,battistiUniversalityPseudogapEmergent2017}, Nickelates~\cite{Li2019, Wu2020}, and Iridates~\cite{Kim2009, Cao2018} host a even rich variety of novel physical behaviors beyond standard lore, including strange metal behavior~\cite{seiboldStrangeMetalBehaviour2021}, and superconductivity~\cite{leeDopingMottInsulator2006,kondoDirectEvidenceCompetition2009,keimerQuantumMatterHightemperature2015} in association with a pseudogap phase~\cite{stanescuPseudogapDopedMott2003,normanPseudogapFriendFoe2005,kohsakaVisualizingEmergencePseudogap2012,tuEvolutionPairingOrders2019} that does not even host a complete Fermi surface.
It is therefore of utmost importance to understand not only the static structure but also the slow dynamics of Mott insulators and their (self-)doped systems.

Even with intensive investigations on simple models~\cite{yokoyamaVariationalMonteCarloStudies1987,yamajiVariationalMonteCarlo1998,giamarchiPhaseDiagramsTwodimensional1991,whitePhaseSeparationStripe2000,scalapinoNumericalResultsHubbard2001,eichenbergerSuperconductivityAntiferromagnetismTwodimensional2007,hettlerNonlocalDynamicalCorrelations1998,hettlerDynamicalClusterApproximation2000,wietekMottInsulatingStates2021} and real materials~\cite{anisimovBandTheoryMott1991,choiFirstprinciplesTreatmentMott2016,grytsiukNb3Cl8PrototypicalLayered2024}, currently the standard understanding of the Mott insulators~\cite{imadaMetalinsulatorTransitions1998} is limited to the ground state description resulting from perturbing a collection of isolated atoms via kinetic coupling between them.
Such a \textit{position} space description of the ground state, while physically intuitive, lacks direct access to the dynamics of long space-time scale relevant to dynamical transport and other response properties, and thus leaves large space for on-going debate on the nature of the quantum~\cite{sachdevQuantumPhasesPhase2004} and thermal~\cite{mottMetalInsulatorTransition1968,imadaMetalinsulatorTransitions1998} phase transition to metals.
On the other hand, the state-of-the art numerical studies based on dynamical mean-field approximation~\cite{AntoineDynamical1996} lead to pictures intimately tied to coherent quasi-particles of the Fermi liquid and thus limited in its applicability.

The difficulty in obtaining a satisfactory understanding for Mott insulators and their (self-)doped metals is profoundly associated with the limitation of the standard framework~\cite{landautheory1937, kadanoffScaling1966, wilsonRENORMALIZATION1974} for thermodynamical and quantum phases.
This framework, developed by Landau~\cite{landautheory1937} and extended by Kadanoff~\cite{kadanoffScaling1966} and Wilson~\cite{wilsonRENORMALIZATION1974}, targets the most relevant information, namely the order parameter (and their correlation) of systems with spontaneously broken symmetries.
It has thus been extremely successful as \textit{the} ``universal'' paradigm for phases with broken charge, magnetic, or orbital symmetries.
Yet, its usefulness and applicability rapidly diminish for phases without a spontaneous broken symmetry, such as the Mott insulating phase to be demonstrated here.
Particularly, to date no knowledge is available on a universal ``fixed point'' of Mott insulators or around it the ``relevant'' dynamics at long space-time scales.
A more general framework for describing quantum phases of matter beyond broken symmetries is therefore of utmost importance.

\begin{figure*}[ht]
\centering
\includegraphics[width=1\textwidth]{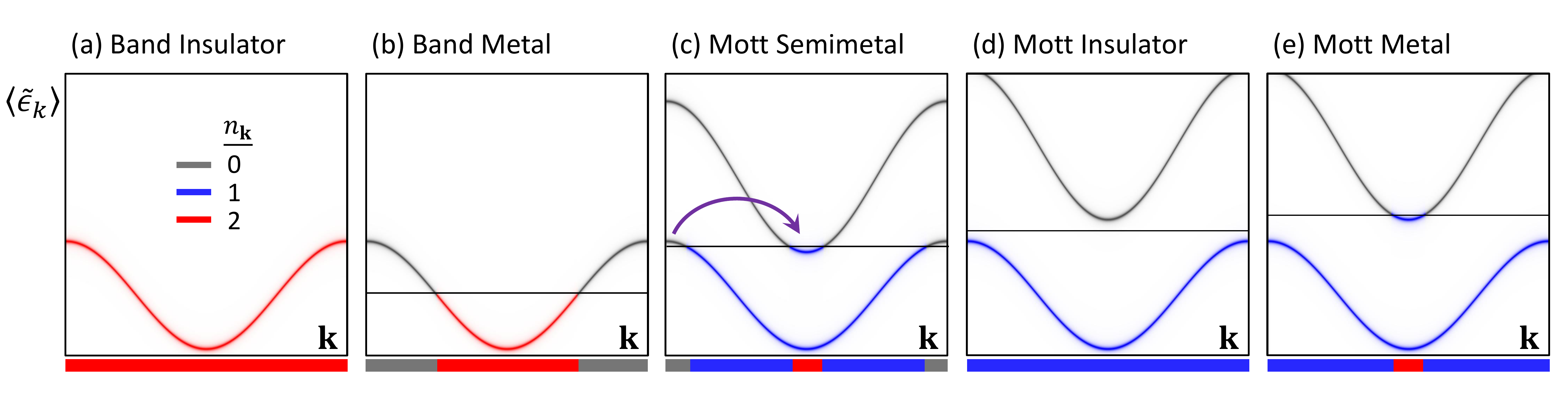}
\vspace{-0.7cm}
\caption{\justifying
Schematics of \textit{eigen-particle} occupation in several quantum phases.
(a) Double occupation (in red) of all momenta in band insulators.
(b) Double occupation within Fermi wavevector ($k<k_F$) in band metals.
(c) Compensating number of zero and double occupations in Mott-semimetals.
(d) Single occupation (in blue) of all momenta in Mott insulators.
(e) Double occupation in few momenta in Mott-metals.
For easier visualization, the dispersions of eigen-particle energy $\braket{\tilde{\epsilon}_k}$ in (c) and (e) are for a reference Mott \textit{insulating} (excited) state of the system, instead of the metallic ground state.
}
\label{fig1}
\vspace{-0.5cm}
\end{figure*}

Here, we address this long-standing problem through a general $N$-body renormalization group framework that offers a generic \textit{dynamics}-based characterization of quantum phases of matter beyond broken symmetries.
The essential ``Mott-ness'' turns out to be characterized by a change of size-scaling of the \textit{effective} intra-\textit{momentum} repulsions between long-lived emergent ``eigen-particles'' that encodes the dynamics of two-body bound states in the high-energy sector.
This directly offers a \textit{universal} characterization of long space-time scale for the corresponding class of Mott insulators through a uniform single occupation of all momenta, and otherwise Mott metals.
This universal description naturally enables the possibility of topological Mott insulators and is straightforward to extend to bosonic Mott systems.
More importantly, the demonstrated general framework offers a generic paradigm to characterize quantum phases of matter through their distinct \textit{dynamics} beyond broken symmetries.

\textit{Continuous Particle-Dressing as a General $N$-Body Renormalization Group Flow} —
We start by acknowledging that quantum phases of matter must have \textit{qualitatively} distinct dynamics of long space-time scales.
For any given Hamiltonian $H[\{c^{\dagger}_{kn\sigma}\}]$ of a lattice translational symmetric system, represented in creation operator $c^{\dagger}_{kn\sigma}$ of momentum $\hbar \mathbf{k}_k$, band index $n$, and spin $\sigma$, such emergent dynamics of long space-time scales can be accessed through systematically absorbing $N$-body dynamical processes of shorter space-time scales into the internal structure of the dressed particles $\tilde{c}^{\dagger}_{kn\sigma}$ via a unitary transformation~\cite{kannoMethodQuasiLinearCanonical1969, wegnerFlowequations1994, white2002numerical},
\begin{equation}
\label{eq:eigenparticle_def}
\begin{aligned}
\tilde{c}^{\dagger}_{kn\sigma} \equiv \mathcal{U}^\dagger c^\dagger_{kn\sigma} \mathcal{U},
\end{aligned}
\end{equation}
such that the corresponding off-diagonal processes are smoothly reduced in the emergent Hamiltonian,
\begin{equation}
\label{eq:H_transform}
\tilde{H}[\{\tilde{c}^{\dagger}_{kn\sigma}\}]\equiv H[\{c^{\dagger}_{kn\sigma}\}]=\mathcal{U}H[\{\tilde{c}^{\dagger}_{kn\sigma}\}]\mathcal{U}^\dagger,
\end{equation}
containing only the more relevant slower dynamics.
Correspondingly, the product states,
\begin{equation}
\label{eq:product_state}
|\Psi\rangle = \prod_j^N \tilde{c}^{\dagger}_{k_j n_j\sigma_j} |0\rangle = \tilde{c}^{\dagger}_{k_1n_1\sigma_1} \tilde{c}^{\dagger}_{k_2n_2\sigma_2} \ldots \tilde{c}^{\dagger}_{k_Nn_N\sigma_N}|0\rangle ,
\end{equation}
of the dressed particles, with $|0\rangle$ denoting the true vacuum with \textit{no} particle,
systematically incorporate the quantum state thermalization~\cite{DeutschETH1991, SrednickiETH2012} up to the finite time scale of the remaining dynamics in $\tilde{H}$.

Upon reaching a diagonal $\tilde{H}$, the fully dressed particles $\tilde{c}^\dagger_{kn\sigma}$ become the infinitely long-lived ``eigen-particles''~\cite{hegg2024universallowtemperaturefluctuationunconventional,kannoMethodQuasiLinearCanonical1969,kannoMethodQuasiLinearCanonical1969a,kannoThermodynamicalFunctionsDilute1969} that give direct access to the slowest dynamics of the system.
Conveniently, since physical dynamics of all space-time scales are fully absorbed, eigen-particles' (1-to-$N$)-body occupations are constants of motion~\cite{zhang2024manipulablecompactmanybodylocalization}.
Consistently, having \textit{fully} incorporated the quantum state thermalization, their product states, $|\Psi\rangle$, form a complete set of $N$-body eigen-states~\cite{10.1143/PTP.41.966}.
This offers a simplest description of many-body eigenstates through \textit{integer} occupation of eigen-particles.

If the above dressing is conducted in a continuous manner with $\mathcal{U}$ only slightly deviating from 1, for example via the Wegner-flow~\cite{wegnerFlowequations1994}, the smooth evolution of the dressed particles $\tilde{c}^{\dagger}_{kn\sigma}$ and their dynamics toward a fully \textit{diagonal} $\tilde{H}[\{\tilde{c}^{\dagger}_{kn\sigma}\}]$ can then be perceived as a generalized renormalization group (RG) flow~\cite{kehreinflow2007} of the \textit{full} $N$-body dynamics toward the longest space-time scale of the system:
First, for systems with a fixed particle number $N$, all possible terms in $\tilde{H}$ together naturally form a closed group during the flow.
Second, the final diagonal $\tilde{H}$ corresponds to a self-similar ``fixed point'', at which further flow for diagonalization returns back to itself.
Third, since dynamics of all space-time scales are absorbed in the dressing of the resulting eigen-particles, $\tilde{c}^{\dagger}_{kn\sigma}$, $\tilde{H}$ contains only the most ``relevant'' dynamics of longest space-time scales.
Therefore, \textit{together with} the particle number and the physical Hilbert space, the final structure of ``relevant'' interactions in $\tilde{H}$ (those \textit{surviving system size scaling}) offers a \textit{complete} $N$-body characterization of all possible quantum phases of matter.

\textit{Band Insulator and Band Metal} —
As a simple example, Fig.~\ref{fig1}(a) illustrates a band insulating state,
\begin{equation}
|\Psi_{\text{BI}}\rangle = \prod_{k\sigma} \tilde{c}^{\dagger}_{k\sigma}|0\rangle,
\end{equation}
in the eigen-particle representation, in which all momentum $k$ of the top valence band are doubly occupied by eigen-particles.
(Without loss of generality, we consider a single valence band for maximal simplicity and clarity.)
In turn, their current fluctuation~\cite{hegg2024universallowtemperaturefluctuationunconventional},
\begin{align}
\tilde{\mathbf{J}}_q = \frac{1}{V}\sum_{k\sigma} \tilde{c}^{\dag}_{k+q,\sigma} \tilde{\mathbf{v}}_{k+\frac{q}{2}} \tilde{c}_{k\sigma}~,
\label{eq:dressed_current}
\end{align}
of wavenumber $q$ is completely disabled by Pauli exclusion principle in the low-energy sector.
Here, the \textit{diagonal} velocity operator of the eigen-particles, $\tilde{\mathbf{v}}_k=\frac{1}{\hbar} \nabla_{k} \tilde{\epsilon}_{k}$, is defined through their diagonal eigen-particle energy \textit{operator} $\tilde{\epsilon}_{k\sigma}$ through $\tilde{c}^\dagger_{k\sigma}\tilde{\epsilon}_{k\sigma}\equiv[\tilde{H},\tilde{c}^\dagger_{k\sigma}]$~\cite{hegg2024universallowtemperaturefluctuationunconventional}.

In contrast, in a band metal state [c.f. Fig.~\ref{fig1}(b)],
\begin{equation}
|\Psi_{\text{BM}}\rangle = \prod_{k<k_F,\sigma} \tilde{c}^{\dagger}_{k\sigma}|0\rangle,
\end{equation}
eigen-particles only doubly occupied the momenta within the Fermi surface bound by $k_\mathrm{F}$.
It thus allows long-wavelength current fluctuation $\tilde{\mathbf{J}}_{q\to 0}$ with nearly zero energy.

As an interesting side note, the abrupt cutoff of the eigen-particle occupation from 2 to 0 at $k_\mathrm{F}$ naturally explains the same for bare particles (highly advocated as the ``quasi-particle weight'' $z$-factor) as a signature of Fermi liquid.
Indeed, under \textit{quantitative} dressing of Pauli-principle-restricted many-body fluctuation in Eq.~\ref{eq:eigenparticle_def}, eigen-particles' preservation of bare-particles' momentum would dictates inheritance of their \textit{qualitative} discontinuity in occupation by bare particles.

\textit{Mott-ness from Strong Local Repulsion} —
To reveal the relevant interaction for Mott-ness at long space-time scales, let's consider the prototypical Hubbard model,
\begin{align}
H=t \sum_{\left< i i^\prime \right> \sigma}c^\dagger_{i\sigma}c_{i^\prime\sigma}+U\sum_i n_{i\uparrow}n_{i\downarrow},
\label{eq:Hubbard_model}
\end{align}
in which the Mott-ness originates from the strong local repulsion of strength $U$, when it dominates over the nearest neighboring kinetic processes of strength $t\ll U$.
For the purpose of revealing the key relevant interaction for Mott-ness, we now proceed to find the canonical transformation that diagonalizes \textit{just} the one-body and two-body terms in such strong interacting limit.

First, decoupling the low-energy sector from the $U$-scaled high-energy one, up to the first order of $t/U$, via $\mathcal{U}\equiv e^A$ with, (c.f. Appendix~\ref{appendix:first_CT})
\begin{align}
A&=\frac{t}{U}\sum_{\left< i i^\prime \right>\sigma} n_{i\bar{\sigma}}c^\dagger_{i\sigma}c_{i\prime\sigma} - h.c.,
\label{eq:A_for_U}
\end{align}
the Hamiltonian represented in the dressed particles, $\hat{H}=\hat{H}^\mathrm{(H)}+\hat{H}^\mathrm{(L)}+\hat{H}^{(n>2)}$, explicitly separates the potential and kinetic motion of the ``doublon'', $\hat{d}^\dagger_i\equiv \hat{c}^\dagger_{i\uparrow}\hat{c}^\dagger_{i\downarrow}$,
\begin{align}
\hat{H}^\mathrm{(H)} = (U+J) \sum_i\hat{d}^\dagger_i \hat{d}_i+ \frac{J}{2} \sum_{\left< i i^\prime \right>}\hat{d}^\dagger_i \hat{d}_{i^\prime},
\label{eq:H_H}
\end{align}
($J\sim 4t^2/U$) in the high-energy sector, from the dynamics of low-energy dressed electrons in a constrained double-occupation-free space, $\hat{H}^\mathrm{(L)}$~\cite{anderson1987resonating, PhysRevB.103.L180502,yildirimKineticsDriven2011, chaoKinetic1977,gros1987antiferromagnetic}.
Here $\hat{H}^{(n>2)}$ describes the beyond-two-body dynamics that emerges from the transformation.
(Beyond the perturbation regime, such decoupling can still be achieved numerically.)

Next, let's bring the one- and two-body terms of $\hat{H}$ into a full diagonal form via another transformation, $\hat{\mathcal{U}}=\exp{(\hat{A})}$, that turns $\hat{c}^\dagger$ into the fully dressed eigen-particles,
$\tilde{c}^\dagger_i\equiv\hat{\mathcal{U}}^\dagger\hat{c}^\dagger_i\hat{\mathcal{U}}$ at the two-body level.
Explicitly,
\begin{align}
\tilde{H}^\mathrm{(H)}=\hat{\mathcal{U}}\hat{H}^\mathrm{(H)}\hat{\mathcal{U}}^\dagger= \sum_i E_i \tilde{d}^\dagger_i\tilde{d}_i,
\label{eq:H_H_diag}
\end{align}
where $E_i=U+J(1-\sum_{\alpha=1}^3 \cos{(\mathbf{q}_i\cdot \mathbf{e}_\alpha)})$ and
\begin{align}
\label{eq:d_diag}
\tilde{d}^\dagger_i&\equiv\hat{\mathcal{U}}^\dagger\hat{d}^\dagger_i\hat{\mathcal{U}}= \frac{1}{\sqrt{V}}\sum_{i^\prime} \hat{d}^\dagger_{i^\prime} e^{i\mathbf{q}_i\cdot \mathbf{r}_{i^\prime}}\\
&=\hat{\mathcal{U}}^\dagger\hat{c}^\dagger_{i\uparrow}\hat{c}^\dagger_{i\downarrow}\hat{\mathcal{U}}=\hat{\mathcal{U}}^\dagger\hat{c}^\dagger_{i\uparrow}\hat{\mathcal{U}}\hat{\mathcal{U}}^\dagger\hat{c}^\dagger_{i\downarrow}\hat{\mathcal{U}}=\tilde{c}^\dagger_{i\uparrow}\tilde{c}^\dagger_{i\downarrow},
\label{eq:d_cc}
\end{align}
with $V$ denoting the system size (the number of sites) and $\mathbf{e}_\alpha$ the unit vectors of the 3-dimensional space.

As expected from the translational symmetry of the system, Eq.~\ref{eq:d_diag} shows that at long space-time scale the eigen-doublons, $\tilde{d}^\dagger_i$, have well-defined momentum $\hbar\mathbf{q}_i$.
The same applies to the constrained particles in the low-energy sector upon diagonalizing $\hat{H}^\mathrm{(L)}$, such that the low-energy contribution (and the one-body component) of $\tilde{c}^\dagger_{i\sigma}$ also has well-defined momentum $\hbar\mathbf{k}_i$ (c.f. Appendix~\ref{appendix:unitary-transformation}).
Therefore, we will substitute $i$ below by the more familiar notation $q$ and $k$ for indexing the fully dressed two- and one-body eigen-particles, $\tilde{d}^\dagger_i$ and $\tilde{c}^\dagger_{i\sigma}$, to better remind the readers that for eigen-particles each $q$ or $k$ indexes a momentum $\mathbf{q}_{q=i}$ or $\mathbf{k}_{k=i}$ that is one-on-one mapped to the a position $\mathbf{r}_i$ of $\hat{d}^\dagger_i$, $\hat{c}^\dagger_{i\sigma}$, and $c^\dagger_{i\sigma}$ according to the gauge choice of $\hat{\mathcal{U}}$ in Eq.~\ref{eq:d_diag}.
Naturally, a consistent periodic boundary condition for $\tilde{d}^\dagger_i$ naturally gives $\mathbf{q}_i=2\mathbf{k}_i$.

Equation~\ref{eq:H_H_diag} and \ref{eq:d_cc} reveals that at long space-time scale, such local repulsion-driven Mott-ness is characterized by the emergence of a ``relevant'' intra-\textit{momentum} repulsion,
\begin{align}
\tilde{H}^\mathrm{(H)}=\sum_q E_q \tilde{d}^\dagger_q\tilde{d}_q = \sum_k E_k \tilde{c}^\dagger_{k\uparrow}\tilde{c}^\dagger_{k\downarrow} \tilde{c}_{k\downarrow} \tilde{c}_{k\uparrow},
\label{eq:IM}
\end{align}
\textit{that scales as} $V^0$ against the system size $V$.
Recall that by default (such as in the $t/U\gg 1$ band metal limit) two-body interactions between particles of well-defined momenta should scale as $V^{-1}$, to reflect the probability for two particles to be in proximity to experience their interaction.
Therefore, a change of scaling of an interaction like this must reflect a \textit{qualitative} change in the underlying correlation.
(Such a change of scaling in the relevant couplings is likely common to all quantum phases as a direct reflection to their characteristic correlations.)

For the specific case here, this change of scaling results from the formation of doublons in the high-energy sector.
Indeed, the dressed particles in the doublon , $\hat{c}^\dagger_{i\uparrow}\hat{c}^\dagger_{i\downarrow}$, are spatially bound to each other, such that the probability for them to experience their mutual interaction is no longer sensitive to the system size, thus the $V^0$-scaling.

Experts in Mott physics might find surprising such emergent intra-\textit{momentum} repulsion in the long space-time scale, given the extreme \textit{spatial} locality of the original strong repulsion.
Note, however, that microscopically Eq.~\ref{eq:IM} only describes the potential and kinetic energy of doublons, rather than two-body interaction between fermions, since it does \textit{not} acts on the low-energy particles at different positions.
The intra-momentum form simply follows eigen-particles' encoding eigen-doublons, $\tilde{d}^\dagger_q$, as equal-momentum pairs, $\tilde{c}^\dagger_{k\uparrow}\tilde{c}^\dagger_{k\downarrow}|_{k=q, (\mathbf{q}_q=2\mathbf{k}_k)}$, in which two high-energy electrons, $\hat{c}^\dagger_{i\uparrow}\hat{c}^\dagger_{i\downarrow}$, are \textit{spatially} bound during propagation (convolution of two momenta of the ``hatted'' particles).
In contrast, pairs of unbound propagating electrons are encoded as $\tilde{c}^\dagger_{k\sigma}\tilde{c}^\dagger_{k^\prime\neq k, \sigma}$.

Naturally, the above consideration on size-scaling would apply to not just the two-body interactions, but all ($n$>1)-body interactions involving doublons.
Upon fully diagonalizing the remaining ($n$>2)-body interactions in $\tilde{H}$ (which preserves the already diagonal one- and two-body structures [Appendix~\ref{appendix:order-by-order}]), the relevant intra-momentum interactions for Mott-ness takes the general form, 
\begin{equation}
\tilde{H}_\mathrm{IM}\equiv \sum_k \tilde{c}^\dagger_{k\uparrow}\tilde{c}^\dagger_{k\downarrow}\tilde{U}_k \tilde{c}_{k\downarrow}\tilde{c}_{k\uparrow},
\label{eq:H_IM}
\end{equation}
where $\tilde{U}_k$, defined through $\tilde{U}_k\tilde{c}^\dagger_{k\uparrow}\tilde{c}^\dagger_{k\downarrow}\equiv[\tilde{H},\tilde{c}^\dagger_{k\uparrow}\tilde{c}^\dagger_{k\downarrow}]$, denotes a diagonal operator [up to ($N$$-$2)-body] for the eigen-doublon energy.

While the above illustration employs the prototypical Hubbard model, the revealed characteristics of the ``stable fixed point'' for Mott-ness (emergence of high-energy two-body bound states and the associated $V^0$-scaled $\tilde{H}_\mathrm{IM}$) represent a \textit{universal} class of Mott systems.
Analogous to universality in phase transitions, the same relevant $\tilde{H}_\mathrm{IM}$ can emerge in long space time scale among a wide variety of Mott systems with \textit{distinct} rapid dynamics in $H$.

\textit{Mott Insulator} —
Having identified the above fixed point (relevant interactions) for Mott-ness in long space-time scale, it is straightforward to find a universal description for this class of Mott insulators.
Specifically, as illustrated in Fig.~\ref{fig1}(d), states with single occupation of all \textit{momenta} by eigen-particles of arbitrary spin $\sigma_k$,
\begin{equation}
\ket{\Psi_\mathrm{MI};\{\sigma_k\}} = \prod_k \tilde{c}^{\dagger}_{k,\sigma_k}\ket{0},
\end{equation}
must be insulating under a $V^0$-scaled $\tilde{H}_\mathrm{IM}$, since long-wavelength current fluctuations $\tilde{\mathbf{J}}_{q\to 0}$ of them would unavoidably results in double occupation of momenta and thus require a \textit{finite} energy of scale $\braket{\tilde{U}_k}$.
(Again, consider a one-band system for simplicity and clarity.)
Intuitively, the remaining spin degree of freedom allows $2^V$ number of insulating states, just as expected from the standard positional space description.

Interesting, this universal ``infrared'' description unifies this representative class of Mott insulators with band insulators at the long space-time scale.
Under the constraints $\tilde{c}^\dagger_{k\uparrow}\tilde{c}^\dagger_{k\downarrow}=0$ of the low-energy Hilbert space, eigen-particles are stuck uniformly in momentum space as in the latter (by $\tilde{c}^\dagger_{k\sigma}\tilde{c}^\dagger_{k\sigma}=0$).
Similarly, in both types of insulators, the uniform occupation in momentum trivially eliminates the $\epsilon_k$-scaled kinetic energy, $\sum_k \epsilon_k = 0$, from the system energy, as intuitively expected for insulators.

\textit{Mott Metal} —
Given the insensitivity of the general RG flow of $\tilde{H}$ to the system's particle number, the identified fixed point for Mott-ness above also directly defines a general class of Mott metals.
Analogous to band metals from doping band insulators (both sharing the same relevant $\tilde{H}$), upon introducing extra electrons or holes to Mott insulating states, as illustrated in Fig~\ref{fig1}(e), the doubly occupied or unoccupied momenta open up small but finite $\tilde{U}_k$-free channels for current fluctuation $\tilde{\mathbf{J}}_q$ of long space-time scale.
Consistently, the $\epsilon_{k\sigma}$-scaled one-body energies no longer vanishes in the system energy.

Nonetheless, owing to the Mott-ness dictated by the same relevant $\tilde{H}_\mathrm{IM}$, Mott metals retain most physical properties of Mott insulators, such as strong charge correlation and large spin entropy at low temperature (c.f. Tab.~\ref{tab1}).
Particularly, the low-energy carrier density for $\tilde{\mathbf{J}}_{q\to 0}$ correspond to the low doping level $x$, rather than the total density,  $1-x$, since most eigen-particles still cannot respond without costing extra $\tilde{H}_\mathrm{IM}$.
These generic properties have been observed in numerous studies~\cite{yokoyamaVariationalMonteCarloStudies1987,yamajiVariationalMonteCarlo1998,giamarchiPhaseDiagramsTwodimensional1991,whitePhaseSeparationStripe2000,scalapinoNumericalResultsHubbard2001} displaying unusual transport~\cite{seiboldStrangeMetalBehaviour2021} and spectroscopic~\cite{stanescuPseudogapDopedMott2003,normanPseudogapFriendFoe2005,kohsakaVisualizingEmergencePseudogap2012} features distinct from the band metals.

\textit{Mott Semimetal} —
Mott metals can also arrive from repopulating eigen-particles in Mott insulators.
As illustrated in Fig.~\ref{fig1}(c), when $\tilde{U}_k$ smoothly reduces to slightly below the dressed bandwidth $\tilde{W}\equiv \max_{k\sigma k^\prime\sigma^\prime}\{\tilde{\epsilon}_{k\sigma} - \tilde{\epsilon}_{k^\prime\sigma^\prime}\}$, relocating eigen-particles from momenta with the highest $\tilde{\epsilon}_{k\sigma}$ to those with the lowest $\tilde{\epsilon}_{k^\prime\sigma^\prime}$ can lower the system energy more than the interaction energy cost.
The ground state thus switches to a Mott semimetal state containing compensating number of eigen-electrons and holes, in perfect analogy to standard band semimetals that are characterized as negative-gap semiconductors.

\begin{table}[t]
\caption{\justifying
Comparison between band metal, Mott semimetal, Mott insulator, and Mott metal, on system-size scaling of $\tilde{U}_k$, relative strength of dressed/bare one-body bandwidth $\tilde{W}$/$W$, itinerant carrier density $\rho_\mathrm{it}$, Fermi surface (FS), and low-temperature entropy.
(See text.)
}
\begin{tabular}{c|c|c|c}
\toprule
Band Metal & Mott Semimetal & Mott Insulator  & Mott Metal \\ \hline
$\langle\tilde{U}_k\rangle \sim V^{-1}$ & $\langle\tilde{U}_k\rangle \sim V^0$ & $\langle\tilde{U}_k\rangle \sim V^0$ & $\langle\tilde{U}_k\rangle \sim V^0$ \\
$\tilde{W} \sim W$ & $\tilde{W} \gtrsim \tilde{U}_k$ & $\tilde{W} < \tilde{U}_k$ & $\tilde{W} < \tilde{U}_k$ \\
$\rho_\mathrm{it}\sim 1$ & $\rho_\mathrm{it}\ll 1$ & $\rho_\mathrm{it} = 0$ &  $\rho_\mathrm{it}\sim x \ll 1$\\
large FS & small FS ? & no FS  & small FS ? \\
low entropy & high entropy & high entropy & high entropy  \\
\hline\hline
\end{tabular}
\label{tab1}
\vspace{-0.4cm}
\end{table}

\textit{On the Hatsugai-Kohmoto model} —
The above derivation explains why recent studies~\cite{hatsugaiExactlySolvableModel1992,phillipsExactTheorySuperconductivity2020,zhaoTopicReviewHatsugaiKohmoto2024,maiNewApproachStrong2024} of Hatsugai-Kohmoto (HK) model recovered many of the characteristics of Mott insulating phase of the Hubbard model~\cite{hubbard1963,hubbard1964,hubbard1964a},such as the interaction-driven metal-insulator transition precisely at half-filling and the violation of Luttinger's theorem~\cite{hatsugaiExactlySolvableModel1992, zhaoTopicReviewHatsugaiKohmoto2024}.
Indeed, Eq.~\ref{eq:H_IM} shows that $\tilde{H}_\mathrm{IM}$ emerges as the infrared limit of strong local repulsion.
The two-body intra-momentum repulsion in the HK model should therefore be understood as a two-body \textit{reduced} $\tilde{H}_\mathrm{IM}$ with $\tilde{U}_k$ replaced by $\braket{\tilde{U}_k}$.
By including this most relevant $V^0$-scaled interaction, the HK model effective imposes \textit{by hand} the key spatial charge correlation for Mott-ness.
Note, however, that $\tilde{H}_\mathrm{IM}$ is only active on the doublons, $\hat{c}^\dagger_{i\uparrow}\hat{c}^\dagger_{i\downarrow}$, as a \textit{whole}.
Thus, expending $\tilde{H}_\mathrm{IM}$ as infinite-range interaction between unbound particles completely obscures the physics.

\textit{Discussion} —
The eigen-particle description presented here has several important implications on theoretical understanding of Mott insulators and metals.
First, given that in Mott insulators the occupied momenta form a compact manifold, the heavily discussed symmetry-protected topological structures in non-interacting systems~\cite{PhysRevLett.100.156401,Rachel_2010,yoshida2014characterization} may be directly inherited here.
That is, in the eigen-particle description, most known categories of topological insulators can have direct counterparts as \textit{``topological Mott insulators''}.

Second, contrary to band metals, Mott metals generally do \textit{not} have a well-defined momentum cutoff of eigen-particles corresponding to the total density, nor do their spin configuration support a Pauli-principle induced systematic suppression of quantum fluctuations.
Therefore, consistent with previous demonstrations~\cite{phillipsMottness2007,stanescu2007theory,rosch2007breakdown,yang2022violation}, generally there is \textit{no} reason to expect the standard Luttinger sum rule.

Third, since the essential Mott-ness is associated with the emergence of $V^0$-scaled intra-momentum repulsion, $\tilde{H}_\mathrm{IM}$, in the long space-time scale, its extension to spin-less bosonic Mott systems is straightforward.
The Mott insulating states for spin-less bosonic are those with uniform \textit{integer} eigen-particle occupation of all the momenta.

Finally, while this study focuses only on the Mott insulting phase, the general $N$-body RG framework can be applied to produce a universal description of any macroscopic quantum phase, as the resulting fixed point provides all relevant $N$-body dynamics that reflects the essential correlations.
In essence, this dynamics-based characterization of quantum phases is more general than the broken symmetry-based Landau paradigm~\cite{landautheory1937} and more informative than the state-of-the-art RG flow~\cite{wilsonRENORMALIZATION1974}.

\textit{Conclusion} —
Using Mott insulators as a prototypical example, we demonstrate a \textit{dynamics}-based characterization of quantum phases of matter through a general $N$-body renormalization group framework.
The essential ``Mott-ness'' turns out to be characterized by a change of size-scaling of the \textit{effective} intra-\textit{momentum} repulsions between long-lived emergent ``eigen-particles'' that encodes the dynamics of two-body bound states in the high-energy sector.
This directly offers a \textit{universal} characterization of long space-time scale for the corresponding class of Mott insulators, through a uniform single occupation of all momenta, and otherwise Mott metals.
This universal description naturally paves ways to topological Mott insulators and is straightforward to extend to bosonic Mott systems.
Finally, this demonstration exemplifies a generic paradigm of characterizing quantum phases of matter through their distinct \textit{dynamics} beyond broken symmetries.

\begin{acknowledgments}
\textit{Acknowledgment} —
We thank P. W. Phillips for useful discussion on the HK model and the symmetry property of the Fermi liquid.
This work is supported by the National Natural Science Foundation of China (NSFC) under Grants No. 12274287 and No. 12042507 and the Innovation Program for Quantum Science and Technology No. 2021ZD0301900.
\end{acknowledgments}

\appendix

\section{Mott Transition to Mott semimetal}

According to Fig.~\ref{fig1}, under a smoothly reducing $\tilde{U}_k/\tilde{W}$, the transition from Mott insulators to Mott semimetals, commonly named ``Mott transition'', should display a continuously growing carrier density according to the amount of repopulated eigen-particles from the reference Mott insulating states, analogous to standard phase transition from band semiconductors to band semimetals.
Indeed, many numerical studies~\cite{furukawa2015quantum,li2021continuous,kiely2024continuous,zhang1993mott} found continuous quantum phase transitions for this class of local-interaction driven Mott insulators.

Note that compared to the remaining slow dynamics in the low-energy sector, the most relevant $\tilde{U}_k$ is of rather high energy scale, so should the energy scale for the emergence of Mott-ness.
This leaves a large space for additional relevant interactions to emerge  below the scale of $\tilde{U}_k$, which manifest themselves through further bifurcations in the general $N$-body RG flows toward different fixed points (all with Mott-ness).
Therefore, the corresponding quantum phase transitions between these Mott systems can sometimes become first-order to reflect the more dramatic change of low-energy correlation, as observed in previous studies~\cite{PhysRevMaterials.3.054605}.
Nonetheless, we consider such first-order nature of transitions unrelated to the continuous nature of quantum insulator-to-metal transition of the \textit{generic} Mott systems having $\tilde{H}_\mathrm{IM}$ as the \textit{only} relevant interaction.

From this perspective, Mott's original proposal~\cite{mottMetalinsulatorTransitionsVO21974} of first-order phase transition is perhaps beyond this class of generic Mott transition (defined above via strong local repulsion).
This is because Mott's proposal is based on the competition between spatial size of bound electron-hole pairs in insulating states and the length scale of screened interaction in metallic states.
If the single occupation of the above $\hat{c}^\dagger_{i\sigma}$ in position are viewed as local dressed electron-hole bound pairs, the extreme locality of their binding would be completely insensitive to the long-range interaction.
On the other hand, for Mott's proposal to apply to the low-density electron and hole carriers in Mott-semimetal states, emergence of additional $V^0$-scaled \textit{inter-momentum} repulsion beyond $\tilde{H}_\mathrm{IM}$ would be necessary to encode the excitonic binding at long space-time scale.

Note that the above conclusion of \textit{default} continuous Mott transition is based on the smoothness of unitary transformation of the full many-body structure.
Presence of non-linearity from extrinsic sources, such as strong coupling to lattice degrees of freedom, can in principle allow a (weak) first-order transition as well.

\section{Unitary Transformation for the Fourier Transform} \label{appendix:fourier}

For systems with translational symmetry, eigen-particles and their bound particles should have well-defined momentum.
Thus, a Fourier transform of the objects, for example from $\hat{c}^\dagger_i$ at position $\mathbf{r}_i$ to their momentum counterparts $\tilde{c}^\dagger_i$ of momentum $\mathbf{k}_i$,
\begin{align}
\label{eq:c_FT}
\tilde{c}^\dagger_i \equiv\hat{\mathcal{U}}^\dagger\hat{c}^\dagger_i\hat{\mathcal{U}}= \sum_{i^\prime} \hat{c}^\dagger_{i^\prime} M_{i^\prime i},
\end{align}
would often automatically diagonalize the corresponding terms in the Hamiltonian, where $M_{i^\prime i}=\frac{1}{\sqrt{V}}e^{i\mathbf{k}_i\cdot \mathbf{r}_{i^\prime}}$ is the matrix element for Fourier transform.
This section gives the general procedure to construct the corresponding unitary transformation $\hat{\mathcal{U}}$.

We first define the Hermitian generator $g$,
\begin{equation}
g=-i\ln(\mathcal{U})=\sum_{ii^\prime} c^{\dagger}_i g_{ii^\prime} c_{i^\prime},
\end{equation}
with coefficient $g_{ii^\prime}$, such that $\mathcal{U}=e^{ig}$ produces the desired unitary transformation Eq.~\ref{eq:c_FT}.
Given that the first and second quantization share the identical algebra when applying a unitary transformation Eq.~\ref{eq:c_FT} via the Baker–Campbell–Hausdorff formula~\cite{baker1901further,hausdorff1906symbolische,campbell1896law}, the matrix $\mathbf{g}$ with coefficients $g_{ii^\prime}$ must be
\begin{equation}
\mathbf{g}=-i\ln(\mathbf{M}^{\dagger}),
\end{equation}
as well, with matrix $\mathbf{M}$ containing elements $M_{i^\prime i}$.
Here we use the principal value definition of the logarithm, where the imaginary part of the lies in the interval $\left(-\pi,\pi\right]$.
In numerical implementation, it is critical to control eigenvalues to ensure consistency of this interval.

Then, given the special property $\mathbf{\mathbf{M}}^{\dagger 4}=\mathbf{I}$, matrix $\mathbf{g}$ can be obtained,
\begin{align*}
\mathbf{g} & =\frac{\pi}{2}\mathbf{G}_{1}+\pi \mathbf{G}_{2}-\frac{\pi}{2}\mathbf{G}_{3},
\end{align*}
through the orthogonal projection matrices,
\begin{align}
\mathbf{G}_{1} & =\frac{1}{4}(\mathbf{I}-i\mathbf{\mathbf{M}}^\dagger-\mathbf{\mathbf{M}}^{\dagger 2}+i\mathbf{\mathbf{M}}^{\dagger 3}),\label{eq:G1}\\
\mathbf{G}_{2} & =\frac{1}{4}(\mathbf{I}-\mathbf{\mathbf{M}}^\dagger+\mathbf{\mathbf{M}}^{\dagger 2}-\mathbf{\mathbf{M}}^{\dagger 3}),\label{eq:G2}\\
\mathbf{G}_{3} & =\frac{1}{4}(\mathbf{I}+i\mathbf{\mathbf{M}}^\dagger-\mathbf{\mathbf{M}}^{\dagger 2}-i\mathbf{\mathbf{M}}^{\dagger 3}),\label{eq:G3}
\end{align}
according to Ref.~\cite{Aristidou2007}.

\section{Order-by-order diagonalization}\label{appendix:order-by-order}
In this section we show that diagonalization via canonical transformation can be performed \textit{order-by-order} in second quantized form.
Specifically, starting with $H$ already in diagonal form up to $M$-body with the rest ($n$>$M$)-body terms not in diagonal form, the coefficients of the existing diagonal ($n$$\leq$$M$)-body terms would be preserved upon further canonical transformation to diagonalize the remaining terms.

To see this, consider the anti-Hermitian generator
\begin{equation*}
    A =\sum \alpha_{1\dots(M+1);1'\dots(M+1)'}  c^{\dagger}_1\dots c^{\dagger}_{(M+1)} c_{(M+1)'} \dots c_{1'},
\end{equation*}
of $\mathcal{U}=e^A$ that diagonalizes the ($M+1$)-body terms, through~\cite{baker1901further,hausdorff1906symbolische,campbell1896law},
\begin{equation}
    e^A H e^{-A}=H+[A,H]+\frac{1}{2!}[A,[A,H]]+\dots~~.
\label{BCH}
\end{equation}
It becomes clear that if $[A,H]$ does not generate normal ordered terms of $M$- or fewer body, the previously already diagonal ($n$$\leq$$M$)-body terms would then be preserved.
This is, in fact, the case according to the following theorem.

\begin{theorem*}[Bound in particle number from commutation]\label{thm:nbody-order}
The resulting normal ordered terms of a commutator, $[A_1,A_2]$, for normal ordered operators $A_1$ and $A_2$, of $n_1$- and $n_2$-body, respectively, can only be of $m$-body with $m\in[\max\{n_1,n_2\}, n_1+n_2-1]$.
\end{theorem*}

One can verify this theorem by carrying out a few simple cases and observe the basic structure in agreement with the theorem.

The preservation of fewer-body terms during diagonalization of more body terms naturally suggests that one can first diagonalize the most relevant one-body and two-body dynamics, before systematically moving on to three-body and more body dynamics.
Since the canonical transformation at $n$-body level is purely algebraic, unrelated of the actual number of particles, $N$, in the system, the diagonalization can be conducted within the much smaller $n$-body Hilbert space than the $N$-body one of the actual system of interest.

\section{Unitary transformation to absorb intersite charge fluctuation: \texorpdfstring{$c^\dagger_{i\sigma} \to \hat{c}^\dagger_{i\sigma}$}{c†ᵢσ → ĉ†ᵢσ}}
\label{appendix:first_CT}
This section provides more detailed description on the first canonical transformation, from $c^\dagger_{i\sigma}$ to $\hat{c}^\dagger_{i\sigma}$, that absorbs the inter-site charge fluctuation, after which doublons emerge as well-defined objects for dynamics of longer time scale.

First, for on-site repulsion $U\gg t$ much stronger than the inter-site hopping $t$ in the Hubbard model, we identify the double occupation of site as the high-energy object of the two-body Hilbert space, and corresponding, two particles occupying different sites are within the low-energy sector.
The desired canonical transformation can then be expressed as $e^A$, with anti-Hermitian operator $A$ of the form,
\begin{equation*}
A=C\sum_{\left< i i^\prime \right>\sigma} (n_{i\bar{\sigma}}c^\dagger_{i\sigma}c_{i\prime\sigma} - h.c.),
\end{equation*}
where coefficient $C$ can be numerically found to fully decouple the above mention high- and low-energy sector within the \textit{two-body} Hilbert space.
Naturally, $C\sim\frac{t}{U}$ in the simple perturbative $U\gg t$ limit.

Notice that this choice of $A$ is only of two-body.
It is therefore much \textit{simpler} than the typical one used to fully decouple the high-energy sector of the $N$-body Hilbert space (containing at least one doublon) from the low-energy one (containing no doublon.)
Nonetheless, for such decoupling in the two-body Hilbert space, or at the two-body level of the second quantized Hamiltonian, this is sufficient.
(This nicely illustrates the simplicity of the above mentioned order-by-order diagonalization and the convenience of our overall framework.)

Represented by the resulting dressed particles, $\hat{c}^\dagger_{i\sigma}\equiv \mathcal{U}^\dagger c^\dagger_{i\sigma}\mathcal{U}$, using Eq.~\ref{BCH}, the Hamiltonian now reads,
\begin{align}
    \hat{H} &= \hat{H}^{(L)} + \hat{H}^{(H)} \nonumber \\
   &= \left[ \begin{array}{l}
    ~~~t\sum_{\left< i i^\prime \right> \sigma}(\hat{c}^\dagger_{i\sigma}\hat{c}_{i^\prime\sigma}(1-\hat{n}_{i'\bar{\sigma}}) +h.c.)\\
    + J \sum_{\left<ii'\right>}(\hat{\mathbf{S}}_i\cdot \hat{\mathbf{S}}_{i'}-\frac{1}{4}\hat{n}_i \hat{n}_{i'})\nonumber\\
    - \frac{J}{4}\sum_{\left<ii'i''\right>}(\hat{c}_{i\sigma}^\dagger \hat{n}_{i'\bar{\sigma}} \hat{c}_{i''\sigma}-\hat{c}_{i\sigma}^\dagger c_{i'\bar{\sigma}}^\dagger \hat{c}_{i'\sigma}\hat{c}_{i''\bar{\sigma}}+h.c.) \end{array} \right]\\
    &+ \left[ (U+J)\sum_{i}\hat{d}_i^\dagger \hat{d}_i + \frac{J}{2}\sum_{\left<ii'\right>} (\hat{d}_i^\dagger \hat{d}_{i'}+h.c.)\right],
\label{H_hat}
\end{align}
where $J=4t^2/U $ in the perturbable limit, $\hat{n}_{i\sigma}\equiv \hat{c}^\dagger_{i\sigma}\hat{c}_{i\sigma}$, $\hat{\mathbf{S}}_i \equiv \frac{1}{2}\sum_{\alpha \beta}\hat{c}^\dagger_{i\alpha}\mathbf{\sigma}_{\alpha \beta}\hat{c}_{i\beta}$, $\hat{d}_i^\dagger\equiv\hat{c}^\dagger_{i\uparrow}\hat{c}^\dagger_{i\downarrow}$, and $\left<ii'i''\right>$ denoting three neighboring sites.

The high-energy sector contains only the dynamics of doublon, $\hat{d}_i^\dagger$, in the two-body Hilbert space.
In contrast, the low-energy sector contains only two particles occupying \textit{different} sites ($i\neq i^\prime$) that experience emergent interactions in the second and third line of Eq.~\ref{H_hat}.
The sectors are now cleanly separated in the two-body level.
In the 3- and ($n$>3)-body Hilbert space, additional couplings are present, which can be further diagonalized order-by-order, if desired.

\section{Unitary transformation toward fix points of long space-time scale: \texorpdfstring{$\hat{c}^\dagger_{i\sigma} \to \tilde{c}^\dagger_{i\sigma}$}{ĉ†ᵢσ → c̃†ᵢσ}} \label{appendix:unitary-transformation}

This section provides more detailed description on the second canonical transformation, from $\hat{c}^\dagger_{i\sigma}$ to $\tilde{c}^\dagger_{i\sigma}$, that absorbs the kinetic propagation of the doublon (in the high-energy sector) and the individual single-occupation constrained particles (in the low-energy sector) at very long space-time scale, toward an eigen-particle representation.

Given the translational symmetry of the system, one expects such kinetic propagation would be naturally absorbed via some sort of Fourier transform, such that the index $i$ would switch from labeling the site location $\mathbf{r}_i$ to labeling the momentum quantum number $\mathbf{k}_i$.
Indeed, applying the general $g_{ii^\prime}$ from Appendix~\ref{appendix:fourier} in defining the anti-Hermitian $\hat{A}=\hat{A}^{(H)}+\hat{A}^{(L)}$ with
\begin{equation}
    \hat{A}^{(H)} = \sum_{\left<ii'\right>}\hat{c}^\dagger_{i\uparrow}\hat{c}^\dagger_{i\downarrow} g_{ii'}\hat{c}_{i^\prime \downarrow}\hat{c}_{i^\prime \uparrow}-h.c.,
\end{equation}
and
\begin{equation}
    \hat{A}^{(L)} = \sum_{\left< i i^\prime \right> \sigma}\hat{c}^\dagger_{i\sigma}g_{ii'}\hat{c}_{i^\prime\sigma}(1-\hat{n}_{i'\bar{\sigma}}) -h.c.,
\end{equation}
one transforms the Hamiltonian to a nearly diagonal form $\tilde{H}=\tilde{H}^{(H)}+\tilde{H}^{(L)}$, with
\begin{equation}
    \tilde{H}^{(H)}=\sum_{i}E_i \tilde{c}^\dagger_{i\uparrow}\tilde{c}^\dagger_{i\downarrow} \tilde{c}_{i\downarrow}\tilde{c}_{i\uparrow},
\label{tilde_H_H}
\end{equation}
and
\begin{align}
    \tilde{H}^{(L)} = \sum_i \left[\epsilon_i (\tilde{c}^\dagger_{i\uparrow}\tilde{c}_{i\uparrow}+\tilde{c}^\dagger_{i\downarrow}\tilde{c}_{i\downarrow})-2\epsilon_i\tilde{c}^\dagger_{i\uparrow}\tilde{c}^\dagger_{i\downarrow} \tilde{c}_{i\downarrow}\tilde{c}_{i\uparrow}\right] + \dots~~,
\label{tilde_H_L}
\end{align}
where $E_i=U+J(1-\sum_{\alpha=1}^3 \cos{(\mathbf{q}_i\cdot \mathbf{e}_\alpha)})$ gives the energy dispersion of the doublon with momentum $\mathbf{q}_i$, and $\epsilon_i$ gives the one-body energy dispersion of the individual electrons of momentum $\mathbf{k}_i$.
Notice that $\epsilon_i$ remains the same as the bare kinetic energy obtained from diagonalizing the one-body terms in Eq.~\ref{eq:Hubbard_model}.
(Recall its preservation during diagonalization for the ($n$>1)-body terms in Appendix~\ref{appendix:order-by-order}.)

Also notice that a $V^0$-scaled intra-momentum terms emerges as well in $\tilde{H}^{(L)}$, to exactly cancel the one-body kinetic contributions $\epsilon_i$, as the electrons as part of the doublon cannot propagate independently, thus their one-body kinetics is thus unrelated to the dispersion of doublons.

Strictly, this second canonical transformation still leave some rather small two-body terms in Eq.~\ref{tilde_H_L}, of order $J\ll\epsilon_i$.
We have verified numerically that a complete diagonalization of these small terms does not qualitatively alter the main structures of Eq.~\ref{tilde_H_L}.

\end{document}